\documentstyle[cltp]{article}
\def\cher{$\check{\rm C}$erenkov }
\begin{document}
\input psfig

\title{The \cher Correlated Timing Detector:\\
Materials, Geometry and Timing Constraints}

\author{D. Aronstein, T. Bergfeld, D. Horton, M. Palmer, M. Selen$^\star$, 
G. Thayer}
\address{Department of Physics, 
University of Illinois at Urbana-Champaign, Urbana, Illinois}
\author{V. Boyer, K. Honscheid}
\address{The Ohio State University, Columbus, Ohio}
\author{H. Kichimi, Y. Sugaya, H. Yamaguchi, Y. Yoshimura}
\address{KEK, Tsukuba, Japan}
\author{S. Kanda, S. Olsen, K. Ueno}
\address{University of Hawaii, Honolulu, Hawaii}
\author{N. Tamura, K. Yoshimura}
\address{Okayama University, Okayama, Japan}
\author{C. Lu, D. Marlow, C. Mindas, E. Prebys}
\address{Princeton University, Princeton, New Jersey}
\author{P. Pomianowski}
\address{Virginia Polytechnical Institute, Blacksburg, Virginia}

\twocolumn[\maketitle\par
\begin{abstract}
The key parameters of \cher Correlated Timing (CCT) detectors are
discussed.  Measurements of radiator geometry, optical properties of
radiator and coupling materials, and photon 
detector timing performance are presented.
\bigskip

\end{abstract}
]

\section{Introduction:}
The CCT concept has been outlined in a previous publication
\cite{nim94} and described in detail in an internal CLEO 
collaboration document\cite{cbx}.  The technique uses the
correlation between photon pathlength and \cher production 
angle to infer this angle by measuring the time taken for
the totally internally reflected \cher photons
to ``bounce'' to the end of the radiator.
The most typical radiator geometry is 
that of a quartz bar having rectangular 
cross-section, typically a few centimeters on a side, and 
having a length of about a meter.  This is very similar to
the radiators in the DIRC design of Ratcliff\cite{dirc}.
The key parameters of a CCT system will all be related to 
timing performance.  In practical terms this translates to
photo detector efficiency and transit-time jitter, quality
of the radiator bars in terms of geometry 
and transparency, and our ability to make high quality optical
couplings between bars and detectors.  All of these issues are
addressed below.
\section{Timing Considerations:}
The technique relies exclusively on fast timing information,
hence an important aspect of its evaluation will be to study the 
various available photon detector technologies.  The most obvious
of these, and hence the initial choice, are photomultipliers.

Most CCT geometries under study involve radiators whose 
cross-section have widths of about 4 cm, and thicknesses varying
between 1 and 4 cm.  A standard 2" photomultiplier provides an
acceptable match to this. For example, the 12 stage Hamamatsu 
H2431 has an active photocathode diameter of 4.6cm, which will
cover 93\% (100\%) of the end of a $4 \times 4$ cm$^2$ 
($4 \times 2$ cm$^2$) bar.
This particular tube, which was used extensively in our beam-test
experiments\cite{kichimi}, is also one of the fastest available 2" 
photomultiplier tubes, having a single photoelectron 
timing jitter ($\sigma_{1pe}$) of about 160ps.
Even this seemingly outstanding performance may become
a limiting factor in a real CCT device.  For a relativistic 
charged particle normally incident on a quartz bar (n=1.4)
at a distance $z=1$ meter from its end, the angular 
resolution of the \cher angle per unit of timing uncertainty 
($d\theta/dt$) is about 150 mr/ns, degrading to about 550 mr/ns for
tracks incident at 25 degrees from normal incidence\cite{zdepencence}.
If we set
15 mr resolution as a conservative goal (this would provide
$>2\sigma~ K-\pi$ separation up to about 2 GeV/c momentum using
CCT alone\cite{addtof}) we see that we will need timing resolution 
of 100 ps (27 ps) at $0^\circ$ ($25^\circ$) incidence.

It is clear that if we use conventional photomultipliers we 
will have to rely on photon statistics to
obtain the required timing performance.  We would expect the 
resolution to scale roughly as $\sigma_{1pe}/\sqrt{N_{prompt}}$, 
where $N_{prompt}$ 
is the number of photoelectrons detected in a time window 
starting at the nominal photon arrival time, having a width 
small compared to $\sigma_{1pe}$\cite{prompt}.  Clearly, both 
phototube jitter and photon yield are critical parameters.

We have measured the transit-time spread of the 1", 10 stage Hamamatsu 
H5321 photomultiplier, the fastest conventional tube commercially available
at the time of our test.  Our experimental setup was similar to that used
by Kuhlen\cite{kuhlen}, and is not described here.  We measure the standard
deviation of the single-photon transit time to be 
$\sigma_{1pe} = 82\pm 25$ ps, 
consistent with Hamamatsu's claim of 160 ps FWHM.  The timing distribution 
is observed to have a tail on the high-time side, consistent with expectation
\cite{kuhlen}.  
Although the performance of both the 1" tube tested in our
laboratory and the 2" tube used in our beam test were sufficient for the
CCT prototypes built so far, the availability of a photon detector with
much smaller jitter would let us trade timing performance for photon yield.
This in turn would make thinner radiators a possibility, or perhaps even
the use of UV plastic as a replacement for quartz.  Photon yield is discussed
in the following section.
\section{Optical Transmission Studies:}
The number of \cher photons produced by a charged particle 
traversing a radiator depends only on the speed of the particle, and
the index of refraction of the material, which is fairly constant at 
around $n = 1.47$ for most solid radiators. The parameters that will
ultimately determine the number of detected photoelectrons will 
therefore be those
that describe the transport of photons from their production point to
the photodetectors.  Most notable of these is the optical 
transparency of the radiator material and any coupling compounds
used. 

In this section we describe measurements of transmission spectra performed
on three possible radiator materials (fused silica quartz and two kinds
of plastic), as well as several optical coupling compounds.  
Radiation length measurements are presented in another paper\cite{kichimi}.

The transmission vs. wavelength studies were performed using 
McPherson 218 0.3m scanning monochromator illuminated by a deuterium
lamp.  The sample under test was mounted at the exit port of the 
monochromator, and the transmitted light intensity was measured by a
photomultiplier tube feeding its anode current to a precision electrometer.
The wavelength scan and data acquisition was controlled by a PC with
a custom-built interface.  Anode-current versus wavelength 
spectra were accumulated for each of the materials under study, and each was 
normalized to a ``no sample'' spectrum to yield the transmission curves
presented here.  In each case we look for the wavelength at which the material
transmission ``turns on'', (defined as the 50\% transmission point), and
the width of the turn-on, (defined as the change in wavelength between 10\% and 
90\% transmission).  

\begin{figure}
\centerline{\psfig{file=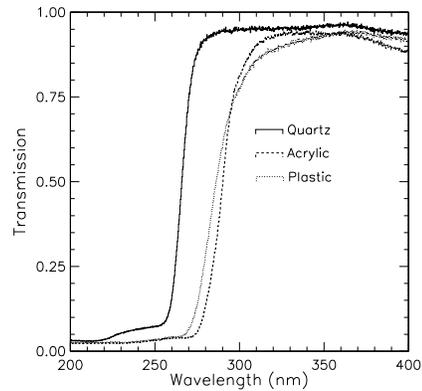,bbllx=92bp,bblly=176bp,bburx=530bp,bbury=626bp,height=5cm}}
\caption{Transmission spectra for quartz (solid) acrylic (dashed) and plastic (dotted) radiators.}
\label{transmission}
\end{figure}

Three radiator materials were tested: 
(i) Nippon fused silica quartz machined into a $4\times 4\times 50$~cm$^3$ bar by
Surface Finishes Inc, (ii) A $2\times 2\times 0.5$~cm$^3$ piece of Mitsubishi 
Acrylite 000 provided by KEK, and (iii) A $2\times 2\times 0.5$~cm$^3$ piece of 
Kuraray plastic LightGuide-S provided by KEK.  The transmission spectra of these
materials is shown in Figure \ref{transmission}.  The quartz is clearly superior,
turning on at a lower wavelength with a sharper edge than either of the organic
materials.
\smallskip

We also tested four possible optical coupling compounds: 
(i) Oken 6262A grease\cite{grease}, provided by KEK, 
(ii) Dow Corning Q2-3067 grease, 
(iii) General Electric Viscasil grease, and 
(iv) Dupont Krytox oil.
The results are summarized in Table 1.
\begin{table}
\begin{center}
\caption{Transmission properties of three possible CCT radiators
and four optical coupling compounds}
\vskip 2 mm
\begin{tabular}{|c|c|c|}
\hline
Material &	$\lambda_{50\%}$ (nm) &	$\Delta\lambda_{10-90}$ (nm)\\
\hline
Quartz	&	$265\pm 3$   &		$15\pm 3$     \\
Acrylic	&	$290\pm 5$   &		$25\pm 5$   \\
Plastic	&	$285\pm 5$   &		$35\pm 10$\\
\hline
\multicolumn{1}{|c}{Coupling Compound} & \multicolumn{2}{c|}{}\\
\hline 
Oken	 &	$280\pm 5$   &		$25\pm 5$ \\
Corning	 &	$275\pm 5$   &		$20\pm 5$   \\
Viscasil &	$195\pm 10$  &		$25\pm 10$    \\
Krytox	 &	$< 150$	     &		--              \\
\hline
\end{tabular}
\end{center}
\label{transprop}
\end{table}
The Oken and Corning compounds are quite similar in performance,
whereas the Viscasil has a significantly lower wavelength cutoff.  Remarkably,
the cutoff of the Krytox oil was below the reach of our apparatus. This 
material is a fluorinated lubricant developed by Dupont, and has
other interesting properties\cite{dupont}.
\section{Quartz Quality Studies:}
The performance of a CCT detector is determined by the time resolution
of the photon detector and by the quality of the quartz radiator. These
two parameters will also set the total cost of the detector.
The price of the quartz is controlled by its UV transparency and by
the quality of the surface finish. 
The second point is closely related to the geometry
of the bar. Standard polishing machines limit the bars to lengths less than
approximately 1.20 m and square bars are easier to handle than designs with
a large aspect ratio. 
Parameters relevant for a CCT device include the transparency of the quartz, 
losses due to absorption or imperfections and the surface quality needed to
preserve the angle information. We have devised a series of tests to study these
parameters and 
the results of our investigation are presented in the following
sections.

Only two manufactures were able to produce quartz bars to our specification:
Surface Finishes Inc. (SF) and Zygo Inc. We bought two 50 cm long
bars from SF and one 120 cm long bar from Zygo in order to 
find out if material of sufficient quality can
be obtained. 
Details of the
specification can be found in Table 2.
\begin{table}
\caption{Quartz Bar Specifications} 
\vskip 2 mm
\begin{center}
\begin{tabular}{|c|c|c|}
\hline
 & SF & Zygo \\
\hline
Grade & \multicolumn{2}{|c|}{Semiconductor}\\
Length & 0.5 m & 1.2 m\\
Width & 4.0 cm & 4.0 cm \\
Height & 4.0 cm & 2.0 cm\\
Roughness & & 7 \AA (rms) \\
Flatness & 0.002 mm & \\
Parallelism & 0.02 mm & 0.01\\
Perpendicularity & 90$^o\pm$30''& \\
Edges & beveled & sharp \\
\hline
\end{tabular}
\end{center}
\label{spec}
\end{table}

UV-grade quartz which is transparent down to wavelengths around 150 nm
should be the material of choice for a {\v C}erenkov detector but our
simulation shows that the gain in the number of photons is offset by
the large dispersion in this wavelength region. For this reason and
to reduce the overall costs we have selected semiconductor grade 
quartz.

\begin{figure}
\centerline{\psfig{file=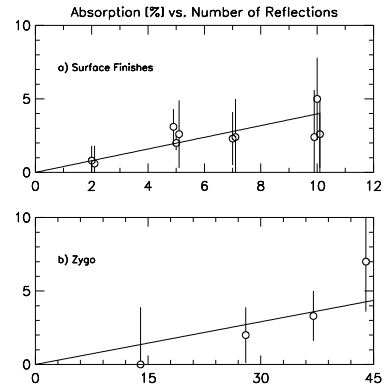,bbllx=92bp,bblly=176bp,bburx=530bp,bbury=626bp,height=5cm}}
\caption{Loss as function of the number of internal reflections for the
Surface Finishes bar (a) and the Zygo bar (b).}
\label{reflection}
\end{figure}

In a CCT detector the {\v C}erenkov photons propagate toward the photon detector via
total internal reflection. 
During this process a small amount of energy travels a short distance as a surface wave
outside the medium. Ideally this is loss free but any surface imperfection 
will change this. With the large number of total internal reflections possible in a CCT device
it is important to study this experimentally.
We have set up a HeNe laser with several Al mirrors that 
give us complete control over the light direction in the quartz bar.
Photodiodes are used to measure the intensity of the incident, reflected and
transmitted laser beam. By changing the angle of incidence we varied the number of
total internal reflections and obtained the results shown in Figure \ref{reflection}.

Fitting to a straight line gives
$0.4 \pm 0.07$\% loss per reflection for the bar from Surface Finishes and for the Zygo
bar we find a 
$0.1\pm 0.03$\% loss per reflection. For 50 bounces this results in a total
loss of 5\% (20\%) for Zygo (SF).
In our study we have carefully avoided the corners of the bar. 
Should the light hit
a corner, losses are significantly larger: For the Zygo bar we measured 
4\% per reflection, and losses are even 
larger for the bar from Surface Finishes due to the beveled edges.
Semiconductor grade quartz can have some inclusions such as air bubbles. 
As much as 10\% of the light was lost when we directed the laser 
beam on a small air bubble in the Zygo bar.

At a momentum of 2.5 GeV or higher the pion and kaon {\v C}erenkov angles differ by
only a few mr and this small angle difference has to be preserved as the 
{\v C}erenkov photons travel toward the photon detector. This places severe
requirements on the geometry of the quartz bar.
In particular, we were interested in determining if the manufacturers 
could achieve the parallelism and perpendicularity specified.

In a first series of tests we used an optical bench to verify that
the macroscopic (geometrical) properties of both quartz bars were
within our specification. We then searched for small scale variations
in the thickness that could change the photon angles. Using two lenses
we widened the beam of a HeNe laser to a spot size of approximately 1 cm
which we directed on one of the long sides of the quartz bar. Changes
in the interference pattern between light reflected at the near and far
side reveal variations in the thickness of the quartz bar. 
\begin{figure}
\centerline{\psfig{file=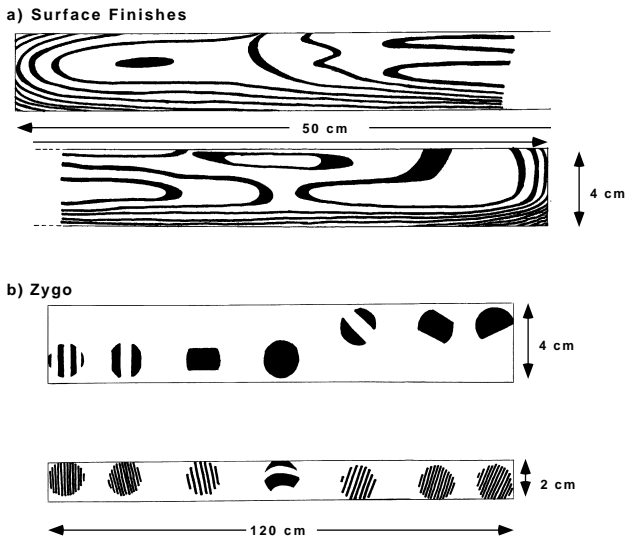,bbllx=42bp,bblly=32bp,bburx=513bp,bbury=434bp,height=7cm}}
\caption{Thickness variation over the length of the bar from Surface Finishes (a). For the
Zygo bar both height and width variations are shown (b).}
\label{interference}
\end{figure}

The results of this study are shown in Figures \ref{interference}a and b for
the Surface Finishes bar and the Zygo bar, respectively. 
The distance between two interference fringes corresponds to a thickness variation of
$\Delta s \; = \;  \lambda/2n \; = \; 0.2 \mu m$\cite{factoroftwo}.

While the wide sides
of the Zygo bar are very flat and parallel, significant curvature is found
for the narrow sides confirming the initial expectation that a large aspect
ratio makes the polishing process more difficult.
For both bars, thickness variations become more pronounced close to 
the corners. Quantitatively, we find a thickness variation of approximately
1-2~$\mu$m for the Surface Finishes bar and better than 
0.5~$\mu$m (2-3~$\mu$m, narrow sides)
for the Zygo bar. Both values are acceptable for a CCT detector.

We gratefully acknowledge the support of the Department of Energy, the
National Science Foundation, and the A. P. Sloan Foundation.
We would also like to express our thanks to Prof. S.Iwata and
Prof. F.Takasaki at KEK for their support of this work.


\begin{thebibliography}{9}
\bibitem{factoroftwo}
The factor of two comes in because
the light passes through the quartz bar twice,
 and $n$ is the reflective index.
\bibitem{prompt}
In other words, the fraction of detected photoelectrons that arrive early enough
to influence the leading edge of the phototube output pulse.
\bibitem{addtof}
The resolution is significantly better if particle time of flight information
is included.
\bibitem{grease}
The term ``grease'' refers to the viscosity of the compound, 
not its chemistry.
\end{thebibliography}
\end{document}